\documentclass[%
 aip,
 jmp,%
 amsmath,amssymb,
 reprint,%
]{revtex4-1}

\usepackage{graphicx}% Include figure files
\usepackage{dcolumn}% Align table columns on decimal point
\usepackage{bm}% bold math
\usepackage{color}
\begin{document}

\preprint{AIP/123-QED}

\title[Dynamic entropy in cancer]{On dynamic network entropy in cancer}% Force line breaks with \\
%\thanks{Footnote to title of article.}

\author{J. West}
 \affiliation{Statistical Cancer Genomics, Paul O'Gorman Building, UCL Cancer Institute, University College London, 72 Huntley Street, London WC1E 6BT, United Kingdom.}%Lines break automatically or can be forced with \\
\author{G. Bianconi}%
\affiliation{ 
 Department of Physics, Northeastern University, Boston, Massachusetts 02115, USA
}%
\author{S. Severini}
\affiliation{%
Department of Computer Science and Department of Physics \& Astronomy, University College London, London WC1E 6BT, United Kingdom.
}%
\author{A.E. Teschendorff}
 \email{a.teschendorff@ucl.ac.uk}
\affiliation{%
Statistical Cancer Genomics, Paul O'Gorman Building, UCL Cancer Institute, University College London, 72 Huntley Street, London WC1E 6BT, United Kingdom.
}%
\date{\today}% It is always \today, today,

\begin{abstract}
The cellular phenotype is described by a complex network of molecular interactions. Elucidating network properties that distinguish disease from the healthy cellular state is therefore of critical importance for gaining systems-level insights into disease mechanisms and ultimately for developing improved therapies. By integrating gene expression data with a protein interaction network to induce a stochastic dynamics on the network, we here demonstrate that cancer cells are characterised by an increase in the dynamic network entropy, compared to cells of normal physiology. Using a fundamental relation between the macroscopic resilience of a dynamical system and the uncertainty (entropy) in the underlying microscopic processes, we argue that cancer cells will be more robust to random gene perturbations. In addition, we formally demonstrate that gene expression differences between normal and cancer tissue are anticorrelated with local dynamic entropy changes, thus providing a systemic link between gene expression changes at the nodes and their local network dynamics. In particular, we also find that genes which drive cell-proliferation in cancer cells and which often encode oncogenes are associated with reductions in the dynamic network entropy. In summary, our results support the view that the observed increased robustness of cancer cells to perturbation and therapy may be due to an increase in the dynamic network entropy that allows cells to adapt to the new cellular stresses. Conversely, genes that exhibit local flux entropy decreases in cancer may render cancer cells more susceptible to targeted intervention and may therefore represent promising drug targets.
\end{abstract}

%Use showkeys class option if keyword
\maketitle

\section{\label{sec:level1}Introduction}
Cancer cells differ from normal cells in terms of a complex landscape of genomic mutations (more generally aberrations), which at a systems-level cause a fundamental dynamic rewiring of the cellular interaction network, ultimately impairing normal cell physiology and allowing cells to acquire key cancer hallmarks \cite{Hanahan2011}. However, the precise systems-level principles underlying the cancer phenotype remain to be elucidated, which represents a key challenge, not only for achieving a deeper understanding of cancer biology but also for identifying novel drug targets \cite{Califano2011}.\\
Ideal drug targets represent genes that upon drug intervention lead to the demise of the cancer cells, while at the same time not compromising the physiology of the normal healthy cells. Given that cellular function is determined by a complex network of protein interactions, one would wish to intervene at points in the network which lead to a functional break-down in cancer cells without affecting their normal cell counterparts. In the parlance of network theory, one seeks to identify nodes in the cancer network which are integral to the robustness of the cancer network, but which are not so (or less so) for the normal healthy network. Thus, it is of interest to explore network properties that may allow us to identify such nodes.\\
In this work, we explore the role of dynamic network entropy in cancer. Our motivation to focus on dynamic network entropy derives from a fluctuation theorem of dynamical systems theory \cite{Demetrius2004} which asserts that the macroscopic resilience of a system, $\mathcal{R}$, is correlated to the level of uncertainty or entropy (disorder), $\mathcal{S}$, of the underlying microsopic dynamical processes that take place within that system. More precisely, the theorem states that the following inequality must hold
\begin{equation}
\Delta\mathcal{S}\Delta\mathcal{R} > 0
\end{equation}
where $\Delta\mathcal{R}$ and $\Delta\mathcal{S}$ represent respectively the changes to the robustness and entropy of the system \cite{Demetrius2004,Demetrius2005}. This inequality holds quite generally, including the case of a stochastic dynamics defined on an underlying network \cite{Demetrius2004,Demetrius2005}. In \cite{Manke2005,Manke2006} this theorem was subsequently applied to protein interaction networks in {\it yeast} and {\it c.elegans}, and it was demonstrated that those genes that contribute most to the network entropy are more likely to be essential genes for the organism. However, in these studies, the stochastic matrix defining the dynamics on the network, and hence the dynamic entropy, was purely topological, i.e the dynamics and entropy were completely specified by the underlying network topology.\\
In order to explore the role of dynamic network entropy in cancer, we use static gene expression data from representative samples of normal and cancer tissue to approximate a stochastic dynamics on a human protein interaction network. Thus, the dynamics refers to the random walk generated by a stochastic matrix on the network, and not to an underlying temporal dynamics, as time course data for individual cancer patients is not available. As we shall see, the dynamics on the network is not entirely specified by the network topology. In fact, we assume that the network topology is unchanged between the normal and cancer phenotypes, but allow the dynamics, defining the weights in the network, to be dependent on the phenotype. Equivalently, we view the protein interaction network as providing only a backbone topological structure as to which interactions are allowed, and use the phenotype-specific gene expression data (and specifically, the correlations in gene expression over the disease phenotype) to modulate and approximate the interaction probabilities. Using this perspective, cancer cells differ from normal cells due to a differential dynamics on the same underlying network.\\
Therefore, our approach is based on two key concepts. First, the integration of gene expression data with protein interaction networks to yield integrated weighted networks, a methodological approach which has already proved fruitful in a variety of different applications within the cancer genomics field \cite{Tuck2006,Pujana2007,Platzer2007,Ulitsky2007,Chuang2007,Milanesi2009,Taylor2009,Hudson2009,Nibbe2010,Yao2010,Komurov2010,Komurov2010bmc,Teschendorff2010bmc,Schramm2010,Vazquez2010}. Secondly, we use the recent notion of ``differential networks'', which attempts to better characterise disease phenotypes by studying the changes in the interaction patterns of these networks \cite{Taylor2009,Hudson2009,Teschendorff2010bmc,Ideker2010,Califano2011,Ideker2012}, as opposed to merely analysing the changes in mean levels of some molecular quantity (e.g gene expression). As demonstrated by several studies \cite{Taylor2009,Hudson2009,Teschendorff2010bmc}, differential networks can identify important gene modules implicated in cancer and also provide critical novel biological insights not obtainable using other approaches. This differential network strategy has recently received further impetus from studies of differential epistasis mapping in yeast, demonstrating that differential interactions may hold the key to understanding the systems-level responses of cells to exogenous and endogenous perturbations, including those present in cancer cells \cite{Ideker2010,Califano2011}.\\
This work is organized as follows. We begin by reviewing the notion of local dynamic network entropy, as constructed from an integrated gene expression protein interaction network, and which was introduced by us previously \cite{Teschendorff2010bmc}. We next extend this notion of local dynamic entropy to a non-local/global one, i.e for extended subnetworks. We then explore the local and non-local dynamic entropy changes associated with the cancer cell phenotype. We point out that previously we had only compared local entropy measures between metastatic and non-metastatic breast cancer \cite{Teschendorff2010bmc}, but not between cancer and normal tissue, as data sets containing larger normal sample collection were less readily available. We also explore the relation between local differential entropy and differential expression. Finally, using the above entropy-robustness theorem, we discuss the potential implications of our findings for devising novel cancer therapies with a view to future studies that will attempt to integrate drug sensitivity data with multi-dimensional (mutational, copy-number, epigenetic and transcriptomic) tumour profiles.

\section{Methods}

\subsection*{The protein interaction network (PIN)}
We downloaded the complete human protein interaction network from Pathway Commons ({\it www.pathwaycommons.org}) (Jan.2011) \cite{Cerami2011}, which brings together protein interactions from several distinct sources. We then built a reduced protein interaction network from integrating the following sources: the Human Protein Reference Database (HPRD) \cite{Prasad2009}, the National Cancer Institute Nature Pathway Interaction Database (NCI-PID) ({\it pid.nci.nih.gov}), the Interactome (Intact) {\it http://www.ebi.ac.uk/intact/} and the Molecular Interaction Database (MINT) {\it http://mint.bio.uniroma2.it/mint/}. Protein interactions in this network include physical stable interactions such as those defining protein complexes, as well as transient interactions such as post-translational modifications and enzymatic reactions found in signal transduction pathways, including 20 highly curated immune and cancer signaling pathways from NetPath ({\it www.netpath.org}) \cite{Kandasamy2010}. We focused on non-redundant interactions, only included nodes with an Entrez gene ID annotation and focused on the maximally conntected component, resulting in a connected network of 10,720 nodes (unique Entrez IDs) and 152,889 documented interactions. In what follows we refer to this network as the ``PIN''.

\subsection*{Normal and cancer tissue gene expression data sets}
We searched Oncomine \cite{Rhodes2004} for studies which (i) had profiled reasonable numbers of cancer and normal tissue samples (at least $\sim 25$ of each type), and (ii) which had been profiled on an Affymetrix platform. In order to reliably estimate covariance of two genes across a set of samples, at least $\sim 25$ samples are needed. The second criterion reflects the desire to conduct the study on a common platform and Affymetrix arrays are the most widely used. Using the same platform across studies ensured that the integrated mRNA-PIN networks were of similar size. In all cases, the intra-array normalised data was downloaded from GEO ({\it www.ncbi.nlm.nih.gov/geo/}), quantile normalized, and subsequently probes mapping to the same Entrez gene ID were averaged. We then subjected each study that passed these criteria through a quality control step, which involved a Principal Component Analysis (PCA) to check that (iii) the dominant component of variation correlated with cancer/normal status. If not, this indicated to us a more pronounced source of non-biological variation, which would confound our downstream analysis. There were six studies satisfying all three criteria and the tissues profiled included bladder (48 normals and 81 cancers) \cite{Sanchez-Carbayo2006}, lung (49 normals and 58 cancers) \cite{Landi2008}, gastric (31 normals and 38 cancers) \cite{DErrico2009}, pancreas (39 normals and cancers) \cite{Badea2008}, cervix (24 normals and 33 cancers \cite{Scotto2008} and liver (23 normals and 35 cancers) \cite{Wurmbach2007}.

\subsection*{Integrated PIN-mRNA expression networks and the stochastic information flux matrix}
For a given cellular phenotype (i.e. cancer or normal), we build an integrated mRNA-PIN using the same procedure as described in \cite{Teschendorff2010bmc}. Briefly, edge weights in the PIN are defined by a stochastic matrix $p_{ij}$,
\begin{equation}
p_{ij}=\frac{w_{ij}}{\sum_{k\in \mathcal{N}(i)}{w_{ik}}}
\end{equation}
with $\sum_{j\in \mathcal{N}(i)}{p_{ij}}=1$, where $\mathcal{N}(i)$ denotes the neighbors of gene $i$ in the PIN and where $w_{ij}=\frac{1}{2}(1+C_{ij})$ denotes the transformed Pearson correlation coefficient $C_{ij}$ of gene expression between genes $i$ and $j$ across the samples belonging to the given phenotype. This definition of $w_{ij}$ reflects our desire to treat correlations and anti-correlations differently. We also note that we enforce $p_{ij}=0$ whenever $(i,j)$ is not an edge in the PIN. Thus, the integrated mRNA-PINs with the edge weights as defined by $p_{ij}$, can be viewed as approximate models of signal transduction flow (as measured by positive gene-gene correlations in expression) subject to the structural constraint of the PIN. Applying this procedure to the two phenotypes yields two integrated PIN-mRNA networks, one for the cancer phenotype with stochastic matrix $p^{(C)}_{ij}$, and one for the normal phenotype with stochastic matrix $p^{(N)}_{ij}$. It is important to point out that the construction of our stochastic matrix means that the topological degrees of each node remain unchanged between the normal and cancer phenotypes: it is only the weights specifying the random walk on the network which differ between the two phenotypes.\\
It is important to stress that we have approximated signal transduction flux on the PIN by positive correlations in expression between interacting genes. This is obviously a crude approximation and therefore a limitation of this study, however, until other types of matched molecular data (e.g protein expression, phosphorylation and other post-translational modification states) become available on a genome-wide basis, we are restricted to the use of only gene expression data. Some further justification for the use of gene expression correlations to approximate signaling flux over the network will be provided by careful comparison of the local correlations to those which are non-local.

\subsection*{A heat kernel stochastic matrix}
It is clear that the stochastic matrix $p_{ij}$ above defines a (biased) random walk on the network $\mathcal{N}$. One may thus compute an information (or probability) flux between any two nodes $i$ and $j$ in $\mathcal{N}$ \cite{Estrada2005}. In fact, it is clear that the probability flux of moving from $i$ to $j$ over a path of length $L$ is given by $(p^L)_{ij}$. It follows that the total probability flux $E_{ij}$ between $i$ and $j$ is given by
\begin{equation}
E_{ij}=\gamma\sum_{L=1}^{\infty}\alpha_L(p^L)_{ij}
\end{equation}
where $\gamma$ is a normalisation factor and where we have introduced a set of arbitrary weights $\alpha_L$ to allow variable contributions for paths of different lengths. One possibility is to suppress paths of longer lengths using $\alpha_L=1/L!$, which also guarantees convergence of the infinite series \cite{Estrada2005}. Formally, defining $\alpha_L=t^L/L!$, we obtain the stochastic matrix 
\begin{equation}
K_{ij}(t)=\frac{\sum_{L=1}^{\infty}\frac{t^L}{L!}(p^L)_{ij}}{e^t-1}
\end{equation}
where we have introduced a ``temperature'' parameter $t$ \cite{Chung2007}. This stochastic matrix is a modified version of the heat-kernel stochastic matrix \cite{Chung2007} and satisfies 
\begin{equation}
\partial_tK(t)=-K(t)(I-p) + \frac{p-K(t)}{e^t-1}
\end{equation}
where we have suppressed matrix indices and where $I$ denotes the identity matrix. Since $p_{ij},K_{ij}(t)\leq 1\quad\forall i,j,t$, it follows that for sufficiently large temperatures ($t\geq 1$), $K(t)$ approximates a solution of the heat-diffusion equation \cite{Chung2007}
\begin{equation}
\partial_tK(t)\approx -K(t)(I-p)
\end{equation}
Thus, the choice $\alpha=t^L/L!$ leads to a natural interpretation in terms of a discrete approximate diffusion process on a graph \cite{Barrat2008}. This construction is therefore closely related to the heat kernel PageRank algorithm  \cite{Chung2007,Brin1998,Barrat2008}.

\subsection*{The dynamic network entropy}
Given the matrix $K_{ij}$, let $Q$ denote the number of non-zero $K_{ij}$, i.e $Q=\sum_{ij}{I(K_{ij}>0)}$ where $I$ is here the indicator function. We then define a dynamic (or flux) entropy as
\begin{equation}
S_{\mathcal{N}}(t)=-\frac{1}{\log{Q}}\sum_{ij}{K_{ij}(t)\log{K_{ij}(t)}}
\end{equation}
where we have rescaled $K_{ij}(t)$ by $1/n$ in order to ensure that $\sum_{ij}K_{ij}(t)=1$. Note that the entropy defined above can be thought of as a non-equilibrium entropy, since the stationary distribution $\pi_i$ of $K_{ij}$, defined by $\pi_iK_{ij}=\pi_j$, was not included. Our choice to consider this non-equilibrium version is motivated by our desire to avoid biasing the entropic contribution of each node to its topological properties (e.g degree).\\
Suppose now that we consider diffusion/flux over paths of maximum length 1. Then, this leads to $K_{ij}=p_{ij}/n$ where $n$ is the number of nodes in $\mathcal{N}$ (we have set $t=1$ for convenience). This leads to the expression
\begin{eqnarray*}
S^{(1)}_{\mathcal{N}}&=&\frac{1}{\log{Q}}\{-\frac{1}{n}\sum_{ij}{p_{ij}\log{p_{ij}}} + \log{n}\} \\
        &=&\frac{1}{\log{Q}}\{\frac{1}{n}\sum_{i}{S_i\log{k_i}} + \log{n}\}
\end{eqnarray*}
In the above expression, $S_i$ is the local dynamic entropy of node $i$ \cite{Barrat2008,Teschendorff2010bmc},
\begin{equation}
S_i= -\frac{1}{\log{k_i}}\sum_{j\in \mathcal{N}(i)}{p_{ij}\log{p_{ij}}}
\label{eq:LFS}
\end{equation}
where $k_i$ is the degree of node $i$ and the normalisation factor ensures that the maximum attainable entropy is equal to 1, independent of the degree of the node.\\
Next, we can consider flux over paths up to length two, in which case
\begin{equation}
K^{(2)}_{ij}=\frac{p_{ij}+\frac{1}{2}(p^2)_{ij}}{\frac{3}{2}n}
\end{equation}
and the corresponding entropy,
\begin{equation}
S^{(2)}_{\mathcal{N}}=-\frac{1}{\log{Q}}\sum_{ij}{K^{(2)}_{ij}\log{K^{(2)}_{ij}}}
\end{equation}
In principle, we can estimate the dynamic entropy $S^{(h)}$ for paths of arbitrary order $h$. In this case,
\begin{equation}
K_{ij}^{(h)}=\frac{1}{n\sum_{r=1}^{h}\frac{1}{r!}}\left(\sum_{r=1}^{h}\frac{1}{r!}\left(p^{r}\right)_{ij}\right)
\end{equation}
In this work we compute dynamic entropies up to moments of order 5 using the R-package {\it expm}. Not going beyond $h=5$ is justified for two reasons: (i) the most interesting behaviour is found for $h\leq 3$, (ii) the computational cost for $h=5$ is considerable, for instance, estimation of flux entropy and associated sampling variance estimates for a typical data set of 30 samples and $\sim 7500$ nodes at $h=5$ takes at least $\sim 20$ hours on a high-performance quad processor workstation.

\subsection*{Sampling variance using the jackknife}
To estimate the statistical significance of observed differences in entropy between two phenotypes, we decided to use the jackknife procedure \cite{Wu1986}. Briefly, the jackknife procedure removes one sample at a time from the given phenotype and recomputes the desired quantity $S$ (here entropy). Thus, if there are $n$ samples in the given phenotype one obtains $n$ jackknife estimates $(\hat{S}_{J,j}:j=1,...,n)$. A jackknife estimate for the mean $S_{\mu}$ and variance $S_V$ of $S$ is then obtained as
\begin{eqnarray*}
\hat{S}_{\mu}&=&n\hat{S}-(n-1)\langle\hat{S}_{J,j}\rangle_j\\
\hat{S}_V&=&\frac{n-1}{n}\sum_{j=1}^n(\hat{S}_{J,j}-\langle\hat{S}_{J,j}\rangle_j)
\end{eqnarray*}
where $\hat{S}$ is the estimate using all $n$ samples and $\langle\hat{S}_{J,j}\rangle_j=\frac{1}{n}\sum_{j=1}^{n}\hat{S}_{J,j}$. Thus, for two phenotypes ``$N$'' and ``$C$'', we compute the difference $\Delta S_J=\hat{S}^{(C)}_{\mu}-\hat{S}^{(N)}_{\mu}$ and obtain a z-statistic
\begin{equation}
z=\frac{\Delta S_J}{\sigma_J}
\end{equation}
where $\sigma_J=\sqrt{S^{(N)}_V+S^{(C)}_V}$.\\
This jackknife procedure can be applied to the dynamic entropy defined over the network or for each node individually. Note that in the case where we obtain z-statistics for each gene/node, the genes can then be ranked according to the significance of this z-statistic. We also note that by construction the z-statistic should be independent of the degree of the node. In fact, while both the differential entropy $\Delta S_J$ as well as the standard deviation estimate $\sigma_J$ will demonstrate the same degree-dependence, the ratio given by the z-statistic $z=\Delta S_J(k)/\sigma_J(k)$ should be degree independent. We demonstrate this empirically across the six different data sets considered here.\\
It should be pointed out that although bootstrapping provides an alternative to the jackknife, that it is not appropriate here since the resampling with replacement would artifically inflate correlations. Another procedure could be to permute the phenotype labels, so that a given ``permuted'' phenotype contains now a mixture of ``normals'' and ``cancers''. However, because there are big differences in expression between normal and cancer, this procedure would dramatically alter the distribution of correlations within the new permuted phenotypes which would also not yield the correct null distribution. Thus, the jackknife strategy circumvents this difficulty while also avoiding the bias associated with bootstrapping.

\section{Results}

\begin{figure*}[!tpb]%figure1
\centerline{\includegraphics[scale=0.75]{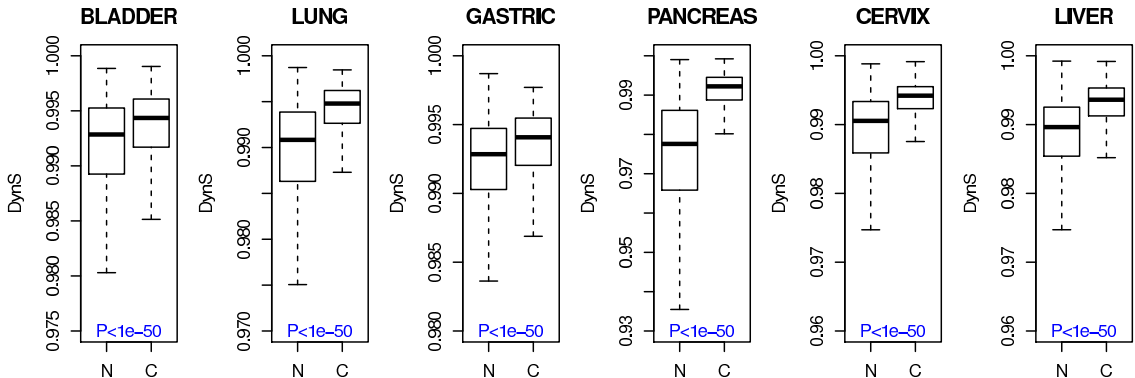}}
\caption{Boxplots of local (i.e. per node) dynamic network entropies (y-axis, DynS) in cancer (C) and normal (N) tissue for all nodes with degree $\geq 10$ ($\sim 3500$ nodes) and across the six different tissue types. P-values are from a one-tailed unpaired Wilcoxon rank sum test. Dynamic entropies have been normalised so that the maximum attainable value is 1. See Supplementary Figure Fig.S2 for the corresponding plot using all nodes with degree $\geq 2$}\label{fig:1}
\end{figure*}

We identified six expression data sets encompassing sufficient numbers of normal and cancer tissue samples and which passed our quality control criteria (Methods). The tissues profiled were bladder, lung, stomach, pancreas, cervix and liver. Integration of these expression data sets with our protein interaction network (PIN) (Methods) yielded sparse weighted networks of approximately 7500 nodes and 98500 edges. The average degree, median degree and diameter of these integrated networks were approximately 26, 8 and 12, respectively. An important assumption underlying any analysis on these integrated networks is that genes which are neighbors in the network are more likely to be correlated at the level of gene expression. While this has been shown for specific data sets (see e.g \cite{Taylor2009}), we verified that it also holds for the integrated mRNA-PIN networks considered here (Fig.S1).

\subsection*{Increased local dynamic entropy is a key hallmark of the cancer cell phenotype}
We previously showed that primary breast cancers that metastasize exhibit an increased dynamic entropy compared to breast cancers that do not spread \cite{Teschendorff2010bmc}. Comparing distinct cancer phenotypes (e.g metastasizing cancers to non-metastasizing) to each other has the advantage that large sample collections are available, thus allowing for more reliable estimates of expression correlations. However, having identified suitable expression data sets encompassing relatively large and balanced numbers of normal and cancer samples, we here sought to determine if the dynamic network entropy also discriminates cancer from its respective normal tissue phenotype. We first compared the local dynamic entropies between normal and cancer, focusing on high-degree nodes (here, nodes with at least 10 neighbours) following the assumption that high degree nodes have higher relevance in cancer \cite{Taylor2009}. Performing this comparison across six different tissue types, using both unpaired and paired non-parametric statistics, clearly confirmed that cancer is indeed characterised by an increased dynamic network entropy (Fig.1, Table~\ref{tab:table1}). Next, we sought to determine if this increased dynamic entropy is also seen if all nodes are included in the analysis. The analogous analysis over all nodes of degree $\geq 2$ (to define the local entropy we need a node to have at least two neighbours) confirmed that dynamic network entropy is increased in cancer (Fig.S2), with the discriminatory power somewhat reduced but still highly significant (Table~\ref{tab:table1}).

%%% table-1
\begin{table}
\caption{\label{tab:table1}Wilcoxon rank sum test statistics comparing the local dynamic entropies (DynS) between normal and cancer, and across the six tissue types. We provide statistics and P-values for the paired (i.e treating the cancer and normal entropies for each gene as dependent variables) Wilcoxon rank sum test. The test-statistics have been normalised to lie between 0 and 1, and thus correspond to an AUC (Area Under receiver operating Curve). AUC values close to 0.5 mean no discrimination, while AUC values closer to 1 indicate a highly significant discrimination between normal and cancer. The corresponding P-values assess the significance of the deviation from 0.5 under a one-tailed test, so that it specifically measures significance of higher entropy in cancer. The top two rows represent the statistics when considering nodes of degree $\geq 10$, while the bottom rows correspond to all nodes for which the entropy can be defined, i.e nodes of degree $\geq 2$.}
\begin{ruledtabular}
\begin{tabular}{ccccccc}
         & BLAD. & LUNG & GAST. & PANC. & CERV. & LIV. \\ \botrule
$k\geq 10$ &  &   &      &      &      &  \\
AUC & 0.75 & 0.92 & 0.69 & 0.97 & 0.88 & 0.88\\
P   & $<10^{-50}$ & $<10^{-50}$ & $<10^{-50}$ & $<10^{-50}$ & $<10^{-50}$ & $<10^{-50}$  \\
$k\geq 2$ &  &   &      &      &      &  \\
AUC & 0.76 & 0.84 & 0.69 & 0.89 & 0.78 & 0.77\\
P   & $<10^{-50}$ & $<10^{-50}$ & $<10^{-50}$ & $<10^{-50}$ & $<10^{-50}$ & $<10^{-50}$  \\
\end{tabular}
\end{ruledtabular}
\end{table}
We observed that the magnitude of differential entropy change was strongly anti-correlated to node degree (Fig.2A). This dependence of dynamic entropy and differential dynamic entropy on the degree of the node was already explored by us previously and reflects an intrinsic bias which needs to be corrected for if meaningful rankings of genes are to be obtained \cite{Teschendorff2010bmc}. In order to correct for this bias, we here devised a statistical framework based on the jackknife to derive z-statistics of differential entropy, which, by construction, would be degree-independent (Methods). Confirming this, we observed that absolute z-statistics did not exhibit a strong anti-correlation with degree, and in fact were on the whole degree-independent (Fig.2B). Supporting our previous result, we also observed that differential entropy z-statistics were significantly higher in cancer compared to normal tissue, independently of tissue type (Fig.S3).
 
\subsection*{Non-local dynamic entropy is increased in cancer, albeit weaker than local dynamic entropy}

\begin{figure*}[!btp]%figure2
\centerline{\includegraphics[scale=0.85]{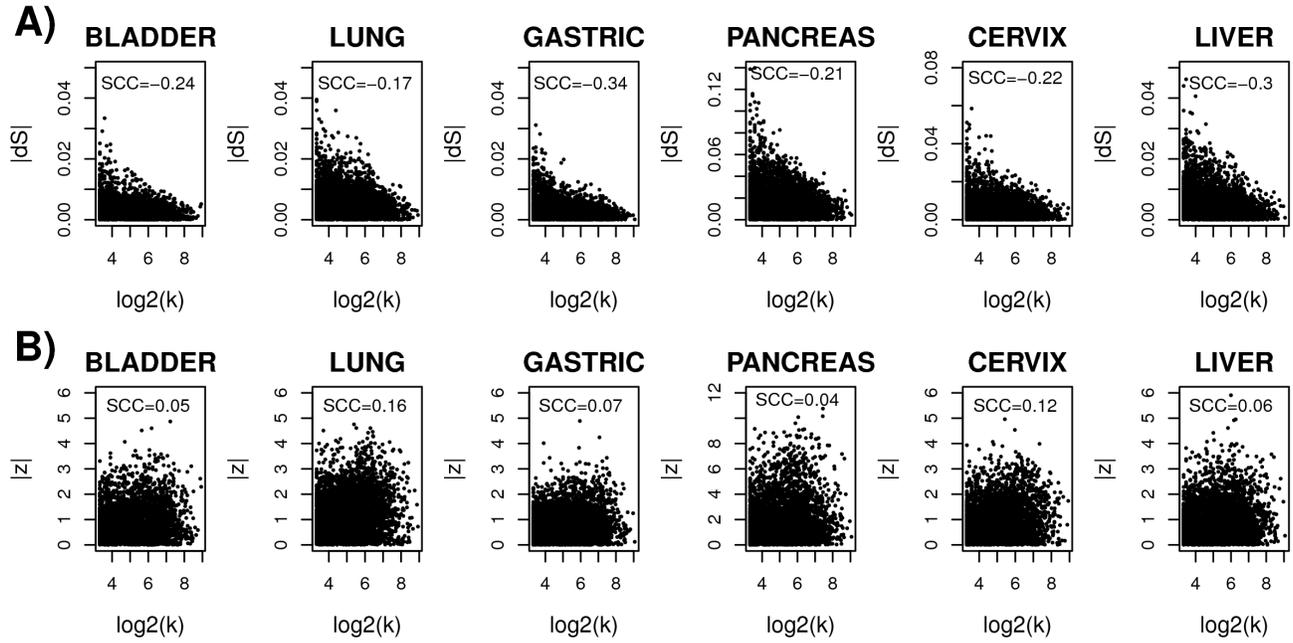}}
\caption{{\bf A)}Scatterplots of absolute differential dynamic entropy changes between normal and cancer (y-axis) against $\log_2(k)$ (x-axis) where $k$ is the degree of the node, for each tissue type. {\bf B)}Scatterplots of the corresponding absolute differential entropy z-statistics (y-axis) against $\log_2(k)$ (x-axis). In both cases, we provide the Spearman rank correlation coefficient.}\label{fig:2}
\end{figure*}

Next, we asked if the higher order dynamic entropy, computed over paths of length larger than 1, are also discriminatory. To this end, we computed for the normal and cancer phenotypes, a higher-order dynamic network entropy 
\begin{equation}
S^{(2)}_{\mathcal{N}}\propto-\sum_{ij}{K^{(2)}_{ij}\log{K^{(2)}_{ij}}}
\end{equation}
where $K^{(2)}_{ij}$ satisfies an approximate diffusion equation over the network allowing for paths of maximum length 2 (Methods). We point out that even when $i$ and $j$ are neighbors, that $K^{(2)}_{ij}$ is not equal to $p_{ij}$, since we allow for alternative signaling paths (of maximum length 2) between genes $i$ and $j$. Thus, this dynamic entropy also takes the well-known redundancy of signaling paths into account \cite{Tieri2010}. For $S^{(2)}$, we also observed a higher dynamic entropy in cancer compared to normal tissue across all tissue types, although this increase was statistically significant only for the four larger studies (Fig.3). We also computed higher order entropies up to paths of maximum length 5. However, as with $S^{(2)}$, higher order dynamic entropies $S^{(k)}, k\geq 3$ generally exhibited reduced discriminatory power, suggesting that the interesting changes associated with dynamic entropy in cancer are localised to neighbors and nearest neighbors in the interaction network.

\begin{figure}[!tpb]%figure3
\centerline{\includegraphics[scale=0.75]{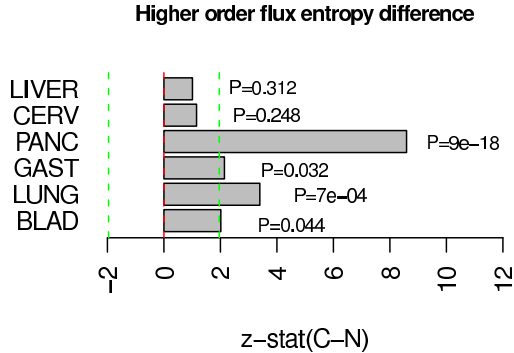}}
\caption{z-statistics of differential non-local dynamic entropy (x-axis) for the six different tissues (y-axis). The dynamic entropy considered here is the $S^{(2)}$ measure which is defined for a stochastic diffusion matrix for maximum path lengths of order 2 (Methods). Positive z-statistics means higher entropy in cancer compared to normal. Green lines indicate the 95$\%$ confidence interval envelope and given P-values are from a normal null distribution centred at zero.}\label{fig:3}
\end{figure}

\subsection*{Differential dynamic entropy and differential expression are anti-correlated}
We argued that if the observed changes in dynamic entropy have a biological basis, that there should be a relationship between the changes in local entropy and gene expression. Specifically, genes which become inactivated in cancer generally exhibit lower expression and this should be reflected as an increased local dynamic entropy around these nodes. Conversely, we hypothesized that genes which become activated in cancer (i.e oncogenes), and which are thus more likely to exhibit higher expression in cancer, would be associated with a lower dynamic entropy since the increased activity of oncogenes is normally associated with activation of specific downstream signal transduction pathways. This means that there is less uncertainty (i.e entropy) along which paths in the network the information flow proceeds. To test this hypothesis, we computed for each gene a regularized t-statistic \cite{Smyth2004} that reflects the degree of differential expression between normal and cancer tissue. Similarly, for each gene we used the previous jackknife procedure to obtain a z-statistic which is a statistical measure of the differential entropy change between the normal and cancer phenotype (Methods).\\

%%% table-2
\begin{table}
\caption{\label{tab:table2}We provide the Pearson Correlation Coefficient (PCC) and P-value (P) between the differential entropy z-statistics $z(dS)$, and the node degree $k$ (top two rows), between the differential entropy z-statistics and the t-statistic of differential expression $t(dE)$ (middle two rows) and finally also the Partial Correlation Coefficient and P-value between $z(dS)$ and $t(dE)$ after adjustment for $k$ (bottom two rows).}
\begin{ruledtabular}
\begin{tabular}{ccccccc}
    & BLAD. & LUNG & GAST. & PANC. & CERV. & LIV. \\ \botrule 
$z(dS)\sim k$ &   &      &       &       &       & \\
PCC & 0.02  & 0.08 & 0.15 & 0.07  & 0.09  & 0.03 \\
P  & 0.76 & 0.06 & 0.09 & 0.02 & 0.22 & 0.72\\ 
$z(dS)\sim t(dE)$ &   &      &       &       &       & \\
PCC & -0.48  & -0.12 & -0.21 & -0.14  & -0.23  & -0.29 \\
P  & 8e-13 & 0.005 & 0.02 & 1e-6 & 0.001 & 6e-5\\ 
$z(dS)\sim t(dE)|k$ &   &      &       &       &       & \\
PCC & -0.49  & -0.14 & -0.23 & -0.15  & -0.23  & -0.29 \\
P  & 6e-15 & 0.001 & 0.009 & 3e-7 & 7e-4 & 3e-5\\ 
\end{tabular}
\end{ruledtabular}
\end{table}

Next, we selected those genes with significant changes in both differential expression and differential dynamic entropy ($P<0.05$). Restricting to these genes, we first verified that differential entropy statistics were not correlated with degree (Table~\ref{tab:table2}). In contrast, differential entropy statistics exhibited a strong anti-correlation with differential expression independently of tissue type, and these anti-correlations remained significant after adjustment for node degree (Table~\ref{tab:table2}). Confirming this analysis further, we observed that genes significantly overexpressed in cancer showed preferential reductions in dynamic entropy compared to genes which were underexpressed, and the associated odds ratios (OR) were statistically significant across all 6 tissue types (Table~\ref{tab:table3}). 

%%% table-3
\begin{table}
\caption{\label{tab:table3}Relation between differential expression and differential dynamical entropy (DynS). The odds ratio (OR) reflects the odds of a gene overexpressed in cancer showing reduced dynamical entropy in cancer, compared to a gene that is underexpressed. The P-value (P) reflects the statistical significance of the odds ratio.}
\begin{ruledtabular}
\begin{tabular}{ccccccc}
    & BLAD. & LUNG & GAST. & PANC. & CERV. & LIV. \\ \botrule 
OR & 6.24  & 3.07 & 2.43 & 2.17  & 3.64  & 2.80 \\
P  & 3e-9 & 0.04 & 0.05 & 0.03 & 0.02 & 0.005\\ 
\end{tabular}
\end{ruledtabular}
\end{table}

\subsection*{Cell-cycle/proliferation genes preferentially associate with a lower dynamic entropy in cancer}
Overexpression of cell-cycle and cell-proliferation genes is a key cancer hallmark with many of these genes representing also candidate drug targets  \cite{Hanahan2011}. Although we have seen that differential entropy changes anti-correlate with differential expression, it is important to check if (1) cell-cycle/proliferation genes are preferentially associated with a reduced dynamic entropy, and (2) whether the anti-correlation between differential entropy and differential expression is driven entirely by cell-cycle genes. To address the first point, we performed a gene set enrichment analysis (GSEA) using the Molecular Signatures Database (MSigDB, \cite{Subramanian2005}) on the top ranked genes, ranked according to the statistics of differential dynamic entropy (separately for increased and reduced entropy). The GSEA analysis showed that genes implicated in the cell-cycle were indeed strongly enriched among genes exhibiting lower dynamic entropy in cancer, but not so among genes exhibiting increases in dynamic entropy (Table~\ref{tab:table4}).\\ 
    
%%% table-4
\begin{table}
\caption{\label{tab:table4} Enrichment analysis of cell-cycle genes among the top 10$\%$ ranked genes exhibiting entropy increases (C$>$N) and decreases (N$>$C) in cancer (C) compared to normal (N) tissue . The enrichment odds ratio (OR) and P-value (P) is from a one-tailed Fisher's exact test. NA=not available due to insufficient number of genes among the top 10$\%$.}
\begin{ruledtabular}
\begin{tabular}{ccccccc}
    & BLAD. & LUNG & GAST. & PANC. & CERV. & LIV. \\ \botrule
DynS(N$>$C) &    &      &      &   &  & \\
OR & 3.92  & 6.07  & 1.35 & NA   & 2.62  & 6.61  \\
P  & 2e-8 & 0.07 & 0.17  & NA  & 0.04 & 4e-11  \\ 
DynS(C$>$N) &   &     &     &   &  & \\
OR & 0.44  & 0.72 & 1.13  & 0.50  & 1.04  & 0.50 \\
P  & 0.99 & 0.93 & 0.36 & 0.99 & 0.46 & 0.99 \\ 
\end{tabular}
\end{ruledtabular}
\end{table}
To address the second point, we repeated the correlation analysis between differential entropy and differential expression, but removing cell-cycle genes prior to the analysis. Importantly, we still observed the anti-correlation between differential entropy and differential expression in 5 of the 6 tissue types (Table S1), indicating that this anticorrelation is a general systemic feature.\\
It could be argued that since tumour expression profiles analyzed here are from the bulk, meaning that the measured expression profiles represent an average over epithelial tumour cells and non-epithelial stromal cells (e.g immune cells), that entropy changes are entirely confounded by changes in the tumour-stromal cell composition. Therefore, it is important to point out here that the enrichment of cell-cycle/proliferation genes among those showing the largest reductions in dynamic entropy, indicates that these differential entropy changes reflect underlying changes in the epithelial tumour cell population, and not changes in the tumour-stromal cell ratio. In other words, the fact that entropy changes can retrieve known tumour cell biology (i.e increased proliferation of tumour cells) shows that interesting tumour cell biology can be extracted from the dynamic entropy.

\section*{Discussion}
In this work we have constructed a dynamic network entropy and have shown that it is increased in cancer compared to normal tissue. Both local and non-local versions of the dynamic entropy were considered, with the local entropy exhibiting the more significant increases. This partly reflects the local nature of expression correlations in the protein interaction network with correlations dropping significantly beyond neighbours and second nearest neighbours (Fig.S4).\\
While dynamic entropy provided a good discrimination between normal and cancer tissue, it is clear that it does not outperform raw gene expression levels, which offer significantly higher classification accuracies \cite{Rhodes2004}. Nevertheless, the fact that dynamic entropy is significantly different between cancer and normal tissue is an important observation, specially in light of the entropy-robustness theorem (Eq.1). Other network measures may also provide equally good discriminators of the cancer phenotype than dynamic entropy. Indeed, the average of the absolute correlations over neighbours of a given node provides an equally good discriminator (Fig.S5), indicating that the loss of local connectivity is a key cancer characteristic. However, the loss of local connectivity (i.e reduced absolute correlations) does correspond to an increase in local dynamic entropy. Therefore, dynamic entropy provides, through inequality $\Delta\mathcal{S}\Delta\mathcal{R} > 0$, a more meaningful framework in which to interpret and understand the systemic changes in gene expression between normal and cancer tissue.\\
It is of importance to discuss (i) what may cause cancer cells to exhibit the observed increase in dynamic entropy and (ii) what it may mean for the cancer phenotype itself. Concerning the first question, one would expect genes that become inactivated in cancer to represent foci of increased dynamic entropy since the inactivation compromises its biological function: at the level of mRNA expression this would manifest itself as reduced expression correlations with its interacting neighbors, but more generally as an increased uncertainty as to which neighbors it may interact with. Conversely, for a gene that is overactivated in cancer its biological function is enhanced, thus confering the cell a selective advantage, which for oncogenes manifests itself as an increased flux of the associated oncogenic pathway. In terms of the local dynamic entropy this increased flux along a particular pathway in the network corresponds to a reduced uncertainty (i.e less dynamic entropy) along which path the information is transferred. In line with these biological expectations we did observe that genes overexpressed in cancer were significantly more likely to exhibit reductions in dynamic entropy than underexpressed genes. Thus, the fact that cancers were characterised globally by an increased dynamic entropy points towards a higher frequency of inactivating over activating alterations in cancer. Intuitively, this makes sense since a random mutation/alteration is more likely to inactivate than activate a gene, and indeed this would be in agreement with recent reports suggesting that most genetic alterations are inactivating and affect tumour suppressors \cite{Wood2007}. We should point out that to formally demonstrate that the increased dynamic entropy is associated with an increased frequency of inactivating alterations (mutations, losses and deletions, promoter DNA methylation) in the tumours analysed here is not possible as matched mutational, copy-number and DNA methylation information is not available for these specific tumours. However, it will be interesting to explore this in the context of the matched multi-dimensional cancer genomic data from the The Cancer Genome Atlas (TCGA) \cite{CGAP2008}.\\
Concerning the second question posed above, we propose that the increased dynamic entropy in cancer could underpin the intrinsic robustness of cancer cells to endogenous and exogeneous perturbations, including therapeutic intervention. This follows directly from the robustness-entropy theorem, $\Delta\mathcal{S}\Delta\mathcal{R} > 0$. However, as we have seen, not all genes exhibit increases in dynamic entropy, with many also exhibiting significant reductions of dynamic entropy in cancer. In particular, we have seen that genes driving cell-proliferation, which are known to be overexpressed in cancer \cite{Rhodes2004}, were preferentially associated with significant reductions in the dynamic entropy. It follows from the above inequality that cancer alterations that are associated with such entropy reductions will make these cells less dynamically robust. Interestingly, this would fit in well with one of the important cancer hallmarks, namely, that of oncogene addiction, whereby cancer cells become overly reliant on a specific oncogenic pathway \cite{Hanahan2011}. Indeed, in cases where the oncogene is druggable, targeting of the oncogene has proved to be an effective drug therapy strategy \cite{Hanahan2011}. Thus, our finding that cell-cycle and cell proliferation genes, which often encode oncogenes, were preferentially associated with reductions in dynamic entropy (and hence reductions in robustness) is consistent with known cancer biology.\\
It follows that the dynamic entropy may be used to identify novel drug targets. As a specific example, we observed that the kinase {\it AURKB} exhibited the largest reductions in dynamic entropy in bladder cancer (Table S2). Importantly, {\it AURKA}, which has already a well established oncogenic role in bladder cancer (see e.g \cite{Park2008}) was also highly ranked (Table S2). Thus, our analysis suggests that the closely related kinase, {\it AURKB}, which has already been implicated as an oncogene and potential drug target in other cancers \cite{Lens2010,Lucena-Araujo2011,Morozova2010}, may also play an equally important role in the pathogenesis of bladder cancer. Given that {\it AURKB} is also druggable (by the drug rebamipide) \cite{Ahmed2011}, this kinase therefore represents an attractive drug target for those bladder cancers that overexpress it. In cases where the oncogene is not directly druggable, we speculate that differential dynamic entropy may be used to identify neighboring druggable targets that also exhibit significant reductions in dynamic entropy. This novel computational strategy could therefore guide non-oncogene addiction based therapeutic strategies that aim to select drug targets within the same oncogenic pathway \cite{Luo2009a,Luo2009b}. Moreover, it has become clear that mutational and copy-number status alone or in combination with gene expression levels are not highly predictive of drug response \cite{Garnett2012,Barretina2012}, hence there is an urgent need for improved in-silico predictors of drug sensitivity. We leave these open and exciting questions for a future bioinformatic study that will analyze matched genomic (mutational, copy-number), epigenomic (DNA methylation), functional (e.g mRNA expression) and drug sensitivity data for large panels of drugs and cancer cell-lines \cite{Garnett2012,Barretina2012}.

\section*{Conclusions}
In summary, in this work we have explored the notion of dynamic network entropy in cancer and have shown that increased dynamic entropy is a key cancer hallmark. Given the correlation between dynamic network entropy and the system's robustness, further investigation of the statistical mechanical principles characterising cancer gene networks is warranted as this may help rationalize the choice of drug targets.

\begin{acknowledgments}
JW is supported by a CoMPLEX PhD studentship. SS is supported by the Royal Society. AET is supported by a Heller Research Fellowship.
\end{acknowledgments}

%\nocite{*}
\bibliographystyle{apsrev4-1}
\bibliography{dynentPREfv}% Produces the bibliography via BibTeX.

\end{document}